\documentclass[prl,twocolumn,showpacs]{revtex4}
\usepackage{amsmath,amssymb,mathrsfs}
\usepackage[applemac]{inputenc}
\usepackage{psfrag}
\usepackage{graphicx}
\usepackage{graphics}
\usepackage{epsfig}
\usepackage{bm}
\usepackage{color}
\usepackage{verbatim,color,ulem}

\begin{document}

\title{Quantum bright solitons in the Bose-Hubbard model \\ 
with site-dependent repulsive interactions}
\author{L. Barbiero$^{1}$}
\email{barbiero@pd.infn.it}
\author{B. A. Malomed$^{2}$}
\email{malomed@post.tau.ac.il}
\author{L. Salasnich$^{1,3}$}
\email{luca.salasnich@unipd.it}
\affiliation{$^{1}$Dipartimento di Fisica e Astronomia 
``Galileo Galilei'' and CNISM,
Universita di Padova, Via Marzolo 8, 35131 Padova, Italy \\
$^{2}$Department of Physical Electronics, School of Electrical Engineering,
Faculty of Engineering, Tel Aviv University, Tel Aviv 69978, Israel\\
$^{3}$Istituto Nazionale di Ottica (INO) del Consiglio Nazionale 
delle Ricerche (CNR), Sezione di Sesto Fiorentino, Via Nello Carrara 1, 
50019 Sesto Fiorentino.}
\date{\today}

\begin{abstract}
We introduce a one-dimensional (1D) spatially inhomogeneous Bose-Hubbard
model (BHM) with the strength of the onsite repulsive interactions growing,
with the discrete coordinate $z_{j}$, as $|z_{j}|^{\alpha }$ with $\alpha >0$
. Recently, the analysis of the mean-field (MF) counterpart of this system
has demonstrated self-trapping of robust unstaggered discrete solitons,
under condition $\alpha >1$. Using the numerically implemented method of the
density matrix renormalization group (DMRG), we demonstrate that, in a
certain range of the interaction, the BHM also features self-trapping of the
ground state into a soliton-like configuration, at $\alpha >1$, and remains
weakly localized at $\alpha <1$. An essential quantum feature found in the
BHM is a residual quasi-constant density of the background surrounding the
soliton-like peak in the ground state, while in the MF limit the
finite-density background is absent. Very strong onsite repulsion eventually
destroys soliton-like states, driving the system, at integer densities, into
the Mott phase with a spatially uniform density
\end{abstract}

\maketitle

\section{Introduction}

The Bose-Hubbard Model (BHM), introduced in 1989 \cite{fisher} as an example
of a system exhibiting a quantum phase transition, has drawn a great deal of
interest -- in particular, due to its experimentally realizability in
ultra-cold Bose gases \cite{bloch}, admitting precise control of interaction
terms \cite{bloch_review}, and availability of probing techniques which can
be applied to this system \cite{greiner, in-situ}. At the same time, BHM is
the ideal platform to study exotic phenomena in reduced dimensions, where
quantum fluctuations can give rise to nontrivial effects \cite%
{giamarchi,cazalilla2011}.

Discrete systems with repulsive interactions, which are described by the BHM
or, in the mean-field (MF) approximation, by the discrete nonlinear Schrö%
dinger equation (DNLSE), give rise, respectively, to quantum \cite%
{quant-dark-sol} or semi-classical \cite{dark-sol} dark solitons. In the
case of weak on-site attractive interactions, the kinetic energy prevents
the collapse, and the system self-traps into bright solitons \cite%
{Salerno,salasnich}. The existence of bright solitons in an attractive Bose
gas was experimentally proved in several experiments \cite{exp-solo1,
exp-solo2, exp-solo3, exp-solo4, exp-solo5}. Furthermore, it was also
demonstrated that repulsive interactions between atoms trapped in an
optical-lattice (OL) potential give rise to gap solitons \cite%
{gap-sol,morsch}, which is another variety of bright modes.

At the mean-field (MF) level, it has been recently shown that, in both
continuum \cite{Barcelona0,Barcelona} and discrete \cite{boris} settings,
repulsive interactions with the strength growing from the center to
periphery faster than $r^{D}$, where $r$ is the distance from the center and
$D$ the spatial dimension, give rise to robust bright soliton-like states.
Because, as is well known \cite{salasnich, haus, drummond, castin, weiss,
weiss2, delande, weiss3}, the MF approximation does not provide for a full
description of physical systems, in this paper we introduce the quantum BHM
with the onsite repulsive interaction growing from the center to periphery.
By means of the density-matrix-renormalization group (DMRG) technique \cite%
{white}, we obtain quasi-exact results for quantum multi-boson bound states,
and compare them to the MF\ prediction \cite{boris} for self-trapped bright
discrete solitons in the zero-temperature model for bosons loaded in a
one-dimensional (1D) OL in the presence of the spatially modulated repulsive
interactions.

The rest of the paper is organized as follows. The model is formulated in
Section II. Numerical results for quantum bound states and their comparison
with the MF counterparts are presented in Section III. It is found that
quantum effects, which the MF approximation cannot grasp, are responsible
for a discrepancy between soliton-like modes in the MF and BHM settings. At
the end of Section III, we study the system in the regime of very strong
self-repulsive interactions. We find that the strong repulsion destroys
self-localized modes, which are replaced by states with a spatially uniform
density. Those states are actually equivalent to well-known Mott insulating
phase in the homogeneous BHM.

\section{The model}

\subsection{The general approach}

We consider a dilute ultracold gas of bosonic atoms confined in the $(x,y)$
plane by the strong transverse harmonic potential,
\begin{equation}
U(x,y)=\left( m\omega _{\bot }^{2}/2\right) \left( x^{2}+y^{2}\right) ,
\label{1}
\end{equation}%
under the simultaneous action of the OL axial potential \cite%
{morsch,book-lattice},
\begin{equation}
V(z)=V_{0}\ \cos ^{2}{(2k_{0}z)}\;.
\end{equation}%
As usual, we focus on the case of the tight transverse confinement, $%
V_{0}\ll \hbar \omega _{\bot }$, which implies a nearly-1D configuration. We
choose the characteristic length of the transverse confinement, $a_{\bot }=%
\sqrt{\hbar /(m\omega _{\bot })}$, and $\hbar \omega _{\bot }$ as length and
energy units, respectively, using accordingly scaled variables below. The
system is described by the quantum-field-theory Hamiltonian,
\begin{eqnarray}
H &=&\int d^{3}\mathbf{r}\ {\psi }^{+}(\mathbf{r})\Big[-{\frac{1}{2}}\nabla
^{2}+U(x,y)+V(z)  \notag \\
&&+\pi g(z)\ {\psi }^{+}(\mathbf{r}){\psi }(\mathbf{r})\Big]{\psi }(\mathbf{r%
})\;,  \label{3dgpe}
\end{eqnarray}%
where ${\psi }(\mathbf{r})$ is the bosonic field operator, and $%
g=2a_{s}(z)/a_{\bot }$, with $a_{s}$ the $s$-wave scattering length of the
inter-atomic interactions \cite{book-bose}.

Unlike previous numerous studies of similar quantum models, here, as said
above, we aim to consider the setting with a $z$-dependent scattering
length, $a_{s}=a_{s}(z)$, which implies $g=g(z)$, as suggested by the recent
analysis of the MF model based on the DNLSE \cite{boris}. The special
inhomogeneity of the nonlinearity strength induces an effective nonlinear
potential \cite{nlattice}, alias a \textit{pseudopotential} \cite{Dong}.
Experimentally the tunability of the magnetic Feshbach resonance (FR) \cite%
{Roati, Pollack} allows the creation of such a spatially inhomogeneous
nonlinearity landscapes by means of properly shaped magnetic fields \cite%
{Boris1}. Furthermore, optically controlled FR \cite{Gora}, as well as
combined magneto-optical control mechanisms \cite{Durr}, make it possible to
create a diverse set of spatial profiles of the self-repulsive nonlinearity.
In particular, the required pattern of the laser-field intensity controlling
the optically induced FR can be \textquotedblleft painted" in space, as
demonstrated in Ref. \cite{Boshier}.

\subsection{The discretization and dimensional reduction}

The presence of the deep OL potential suggests discretization of Hamiltonian
(\ref{3dgpe}) along the $z$ axis. To this end, we use the decomposition in
the general form of \cite{book-lattice}
\begin{equation}
\psi (\mathbf{r})=\sum_{j=1}^{L}{\phi }_{j}(x,y)\ w_{j}(z),  \label{Wannier}
\end{equation}%
where $w_{j}(z)$ is the {Wannier function} maximally localized at the $j$-th
local minimum of the axial periodic potential, and $\phi _{j}\left(
x,y\right) $ are proportional to the ground state of the transverse
potential (\ref{1}),
\begin{equation}
\phi _{j}(x,y)={\frac{1}{\sqrt{\pi }}}\exp {\left[ -\left( {\frac{x^{2}+y^{2}%
}{2}}\right) \right] }\,{b}_{j}\;,  \label{assume}
\end{equation}%
with ${b}_{j}$ representing the bosonic-field operator acting at site $j$,
with $b_{0}=b_{L+1}\equiv 0$. In this work, we consider the case of an even
number $L$ of the lattice sites, see Eq. (\ref{Wannier}).

Next, inserting ansatz (\ref{Wannier}) into Eq. (\ref{3dgpe}), one can
readily derive the effective 1D BHM\ Hamiltonian,
\begin{equation}
H=\sum_{j=1}^{L}\left[ -J\,{b}_{j}^{\dagger }\left( {b}_{j+1}+{b}%
_{j-1}\right) +{\frac{1}{2}}U_{j}{n}_{j}({n}_{j}-1)\right] \;,  \label{ham}
\end{equation}%
where ${n}_{j}={b}_{j}^{\dagger }{b}_{j}$ is the on-site operator of the
number of bosons, while $J$ and $U_{j}$ are the adimensional hopping
(tunneling) amplitude and on-site energy, which are experimentally tunable
via $V_{0}$ and $a_{s}$ \cite{bloch_review}, and are given by
\begin{eqnarray}
J &=&-\int_{-\infty }^{+\infty }w_{j+1}^{\ast }(z)\left[ -{\frac{1}{2}}{%
\frac{\partial ^{2}}{\partial z^{2}}}+V(z)\right] w_{j}(z)\ dz\;, \\
U_{j} &=&\int_{-\infty }^{+\infty }g(z)\ |w_{j}(z)|^{4}\ dz\;.
\end{eqnarray}

In the present model, the hopping energy does not depend on the site number $%
j$, therefore we normalize it to be $J=1$, while, on the contrary to the
standard BHM, the on-site energy $U_{j}$ depends on $j$ through the
inhomogeneous interaction strength, $g(z)$. In particular, choosing
\begin{equation}
g(z)=g_{0}|z|^{\alpha }  \label{alpha}
\end{equation}%
with spatial growth rate $\alpha >0$ [cf. Ref. \cite{Barcelona0}], one has
\begin{equation}
U_{j}=U\ |z_{j}|^{\alpha }\;,  \label{int}
\end{equation}%
where $z_{j}\equiv j-\left( L+1\right) /2$ is the discrete axial coordinate,
and $U_{j}$ attains minimum values
\begin{equation}
U_{\min }\equiv U_{L/2+1}=U_{L/2}=2^{-\alpha }U.  \label{U}
\end{equation}%
at two central sites of the lattice. Thus, in our model the inhomogeneous
on-site energy depends on two parameters, amplitude $U$ and spatial growth
rate $\alpha $.

\subsection{MF (mean field) vs. DMRG (density-matrix renormalization group)}

It is well known that, in 1D configurations, quantum fluctuations, which are
omitted in the mean-field (MF) theory, play a significant role. Thus, in our
1D problem it is relevant to compare MF predictions with those produced by
the DMRG, to conclude in what regimes the MF may give reliable results \cite%
{note}, and to reveal essential quantum features of the ground state beyond
the bounds of the validity of the MF approximation.

We follow the MF approach based on the Glauber coherent state, $|\mathrm{GCS}%
\rangle =|\beta _{1}\rangle \otimes ...\otimes |\beta _{L}\rangle $, where $%
|\beta _{j}\rangle $ is defined so that $b_{j}|\beta _{j}\rangle =\beta
_{j}|\beta _{j}\rangle $, with (complex) eigenvalues $\beta _{j}$ \cite%
{book-luca}. By minimizing energy $\langle \mathrm{GCS}|H|\mathrm{GCS}%
\rangle $ with respect to $\beta _{j}$, one finds that complex numbers $%
\beta _{i}$ satisfy the stationary form of the 1D DNLSE,
\begin{equation}
\mu \,\beta _{i}=\epsilon _{i}\,\beta _{i}-J\,\left( \beta _{i+1}+\beta
_{i-1}\right) +U_{i}|\beta _{i}|^{2}\beta _{i}\;,  \label{dgpe}
\end{equation}%
where $\mu $ is the chemical potential, determined by the total number of
atoms: $N=\sum_{j}|\beta _{j}|^{2}=\sum_{j}\langle \mathrm{GCS}|{n}_{j}|%
\mathrm{GCS}\rangle $. One can solve Eq. (\ref{dgpe}) numerically -- in
particular, with the help of the imaginary-time Crank-Nicolson
predictor-corrector algorithm \cite{sala-numerics}.

The DMRG accounts for the full quantum behavior of 1D lattice systems. As
concerns the search for localized states, in previous works the DMRG has
given strong evidences of bright-soliton states in the homogeneous BHM \cite%
{salasnich}, spin chains \cite{manmana}, and in bosonic models with
nearest-neighbor interactions \cite{das}, including disorder \cite{sacha}.
To produce accurate results by our DMRG-based computations, taking care of
both the interaction with external sites in Hamiltonian (\ref{ham}) and the
needed very large size of the underlying Hilbert space, we utilize open
boundary conditions. We use a number of DMRG states $m$ up to $m=768$,
performing $6$ finite-size sweeps \cite{white}.

\section{Numerical results}

\subsection{Comparison of MF and DMRG results}

\subsubsection{The non-self-trapping case ($\protect\alpha <1$)}

As mentioned above \cite{boris}, the MF predicts the existence of
self-localized states (bright solitons) supported by the spatial modulation (%
\ref{alpha}) with $\alpha >D$ \cite{Barcelona0}. Here, using a range of
parameters similar to \cite{magnons}, we aim to check these predictions in
the BHM\ for $D=1$, by mean of the DMRG technique, and to look for new
effects generated by quantum fluctuations. First, we do this for $\alpha <1$%
, when the self-trapping is not produced by the MF.

\begin{figure}[h]
\centerline{\epsfig{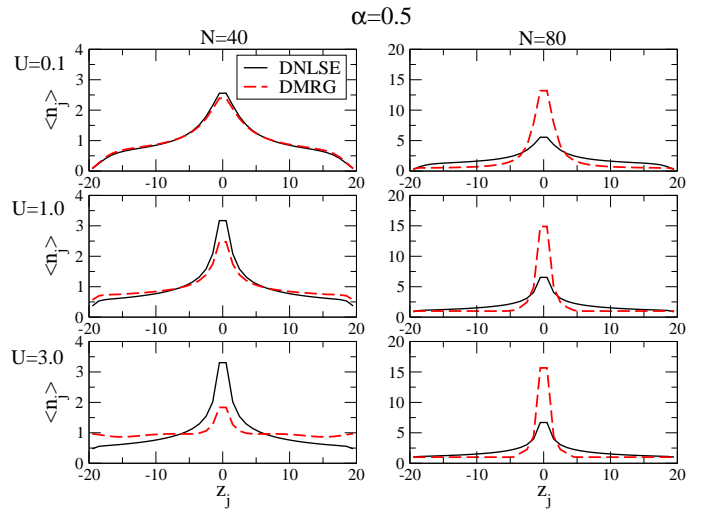}}
\caption{(Color online) DNLSE (mean-field) and DMRG density profiles, $n_{j}$%
, for $\protect\alpha =0.5$, $J=1.0$ and several values of $U$ in the system
with $L=40$ sites and two different values of the numbers of bosons, $N$.}
\label{fig1}
\end{figure}

In Fig. \ref{fig1} we can see that, as predicted by the MF, 
at $\alpha =0.5$, the density profile is not self-trapped 
(i.e., it is not a soliton) for
weak interaction $U$ and average density 
$n\equiv N/L=1$ (the first two panels in 
the left column), the MF and DMRG methods being in very good agreement. This
agreement is explained by the fact that, in certain regimes, quantum
fluctuations are weak and their effect is practically negligible, thus
allowing semi-classical methods, such as the MF approximation, to predict
the correct behavior. Non-self-trapped states are\ also found in the case of
the weak interaction for a larger number of bosons.

On the other hand, the DMRG calculations produce sort of a bright soliton
for stronger interaction (higher $U$), at average 
densities $n=1$ and $n=2$ alike.
More precisely, the DMRG results show that a well-defined peak appears at
the center of the system, surrounded by a nearly flat distribution of the
bosons. Furthermore, while the shape of the density profiles produced by the
MF is actually unchanged, i.e., the shape of the peak depends solely on the
density, but not on the interaction strength, $U$, the DMRG-produced results
do not share this feature. Indeed, for average 
density $n=1$, we observe that, in 
the DMRG states, the fraction of particles located at the center of the
lattice decreases for higher $U$ and, at the same time, the number of bosons
composing the background around the soliton increases. The latter feature is
still more salient for higher average density, i.e. $n=2$. 
This effect can be explained as the possible
approach of the system towards an insulating phase (the Mott phase), in a
region where the interaction strength in Eq. (\ref{int}) exceeds a critical
value of $U$. Further details regarding this point are given below in the
section addressing the Mott phase.

\subsubsection{The self-trapping case ($\protect\alpha >1$)}

The MF theory predicts that spatial inhomogeneity (\ref{int}) with $\alpha
>D $ gives rise to self-trapping into localized states \cite%
{Barcelona0,boris}. To delve deep enough in this regime, we now set $\alpha
=2$.

\begin{figure}[h]
\centerline{\epsfig{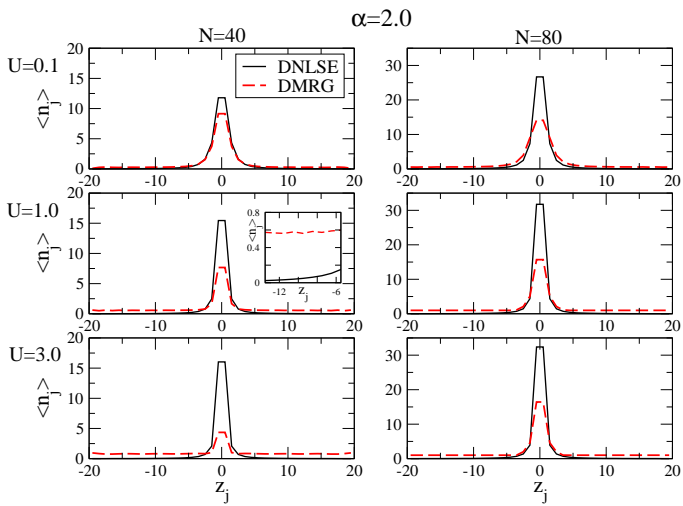}}
\caption{(Color online) DNLSE and DMRG density profiles, $n_{j}$. for $%
\protect\alpha =2.0$, $J=1.0$, and several different values of $U$ in the
system with $L=40$ sites and two different values of the numbers of
particles, $N$. Inset: density profile $n_{j}$ at the first 10 sites of the
chain, for $\protect\alpha =2$, $J=1.0$, $N=40$ and $L=40$}
\label{fig2}
\end{figure}

In Fig. \ref{fig2} it is clearly seen that the MF\ solutions, produced by
the DNLSE, indeed yield well self-trapped states, with vanishing density
around the central peak. Of course, in this case the peak is higher than at $%
\alpha =0.5$, because the corresponding effective interaction is much
stronger in Eq. (\ref{int}).

In the present case, the DMRG results also exhibit, in agreement with their
MF counterparts, self-localized states. Nevertheless, a finite constant
background, which is higher for larger $U$, is again observed around the
quasi-soliton states. Furthermore, due to the stronger interaction, in
comparison to the case of $\alpha =0.5$, the full quantum approach
demonstrates a larger disagreement with MF even at $U=0.1$.

\subsection{Transition from weakly localized to self-trapped states}

\begin{figure}[h]
\centerline{\epsfig{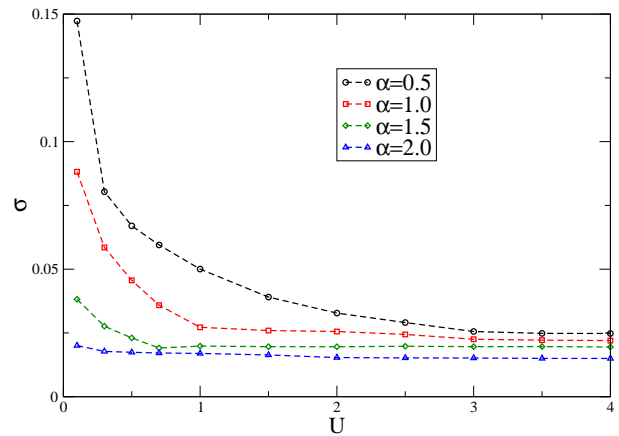}}
\caption{(Color online) DMRG results for the standard deviation $\protect%
\sigma $ computed for $\tilde{L}=10$ [see Eq. (\protect\ref{sigma})], in a
system with $L=40$, $N=40$, $J=1.0$ and different values of $U$ and $\protect%
\alpha $.}
\label{fig5}
\end{figure}

Figures \ref{fig1} and \ref{fig2} clearly demonstrate that the transition
from a weakly localized state to a self-trapped one is driven by the
interaction strength $U$. In particular, the soliton-like state always comes
with a constant finite background, while both the height and the width of
the central peak depend on the interaction strength. In order to clearly
define weakly and strongly localized states, in Fig \ref{fig5} we plot the
standard deviation relative to the external sites of the lattice,
\begin{equation}
\sigma =\sqrt{\frac{\sum_{j}(\langle n_{j}\rangle -\bar{n})^{2}}{\tilde{L}}}
\label{sigma}
\end{equation}%
where $\bar{n}$ is the average value $\langle n_{j}\rangle $ over 
$\tilde{L}$, which is the number of external lattice sites. 
In Fig. \ref{fig5} it is possible to clearly identify,
for all the values of $\alpha $ that we considered, two distinct regimes:
one where the value of $\sigma $ changes for different $U$, meaning that
there is no constant background, and another one, where $\sigma $ is
insensitive to the interaction strength, signaling the appearance of a
practically constant background density profile and thus revealing the
presence of a soliton-like solution existing on top of the background. 
It is worthy to note that, for the strong interaction, a 
self-trapped solution is possible even for $\alpha <1$.

\subsection{Quantum effects}

The purely quantum part in interaction density, $n_{j}(n_{j}-1)$, in the BHM
Hamiltonian (\ref{ham}) is represented by term $-1$ \cite{book-lattice}. To
understand if it is responsible for the appearance of the above-mentioned
nonvanishing background surrounding the soliton peak in the DMRG density
profiles, i.e., if this quantum term is the source of discrepancies between
the MF and DMRG solutions, we have additionally performed the quasi-exact
DMRG-based calculations, but in the \textquotedblleft truncated" form, with $%
\left( n_{j}-1\right) $ replaced by $n_{j}$ in Eq. (\ref{ham}).

\begin{figure}[b]
\centerline{\epsfig{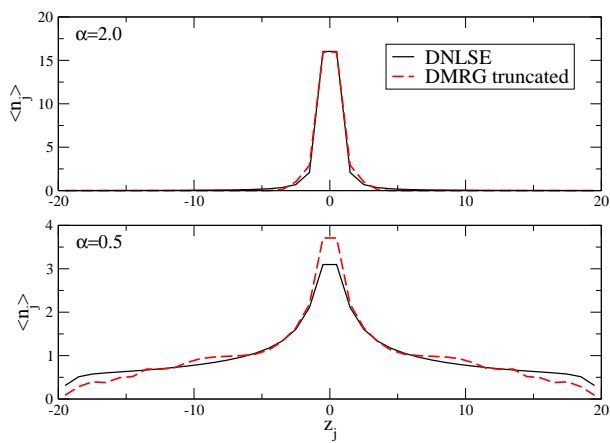}}
\caption{(Color online) The DNLSE and \textquotedblleft truncated-DMRG"
density profiles, $n_{j}$ (without the quantum term $-1$ term in the latter
case, see the text). \textit{Upper panel}: $\protect\alpha =2$, $J=1.0$ and $%
U=3.0$ in the system with $L=40$ sites and $N=40$ particles. \textit{Lower
panel}: $\protect\alpha =0.5$, $J=1.0$ and $U=3.0$ in the system with $L=40$
sites and $N=40$ particles.}
\label{fig3}
\end{figure}

In Fig \ref{fig3} we compare the MF and the truncated-DMRG profiles, for two
different values of $\alpha $, with average density $n=1$ and $U=3.0$. As
shown above in Figs. \ref{fig1} and \ref{fig2}, the MF and normal DMRG
methods are strongly discordant at these values of the parameters. Here we
see that, without the above-mentioned $-1$ term, the finite background
around the peak disappears in the truncated-DMRG state, making it perfectly
similar to the MF counterpart at $\alpha =2$. At $\alpha =0.5$, when
non-self-trapped states are produced by the DNLSE, it also matches well to
the truncated-DMRG counterpart, but in this case small discrepancies are
still visible. Interestingly, the comparison between the lower panel in the
left column of Fig. \ref{fig1} and the lower panel in Fig. \ref{fig3} shows
that the truncated-DMRG results still produce a conspicuous self-trapped
peak, whereas in the case of the full quantum description the peak is very
weak. This comparison demonstrates that truly quantum ingredients of the
system (as a matter of fact, quantum fluctuations) may account for peculiar
effects which are not captured by the MF or nearly-MF description.

\section{Transition from self-trapped states to the Mott phase}

\begin{figure}[t]
\centerline{\epsfig{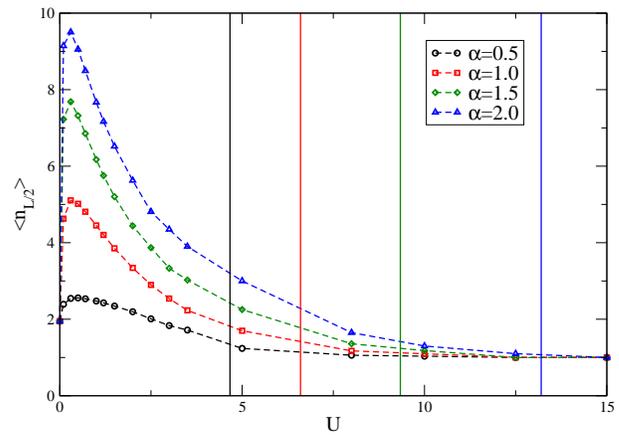}}
\caption{(Color online) DMRG results for the density at the center, $%
\left\langle n_{L/2}\right\rangle $, for $L=40$, $N=40$, $J=1.0$ and
different values of $U$ and $\protect\alpha $. Vertical lines correspond to
the respective critical values for the Mott-superfluid transition in the
finite-size homogeneous system, estimated as per Eq. (\protect\ref{analit}).}
\label{fig4}
\end{figure}

To check if self-trapped states can be always found in the true quantum
system, in Fig. \ref{fig4} we plot the density at the central sites, $%
\left\langle n_{L/2+1}\right\rangle =\left\langle n_{L/2}\right\rangle $
\cite{comment1}, for several values of the interaction strength, $U$, and
spatial growth rate, $\alpha $ in Eq. (\ref{alpha}). More precisely, we aim
to find out how the variation of $U$ modifies the shape of the bosonic cloud
which, as shown before, may be characterized by the density at the two
central sites, where the repulsive interaction is weakest.

Figure \ref{fig4} clearly demonstrates that, at fixed $\alpha $, the height
of the soliton peak in the BHM depends by $U$, on the contrary to the MF
prediction. In particular for weak interactions we find a maximum in $%
\left\langle n_{L/2}\right\rangle $ at a certain value of $U$, which is
nearly the same for all $\alpha $. Then, with the further increase of $U$,
the double occupation become more energetically expensive and the number of
bosons forming the soliton peak decreases, while the boson number increases
in the background surrounding the peak, making the background density closer
to $1$. As seen in Fig. \ref{fig4}, this behavior persists, at all values of
$\alpha $, up to a critical point, $U=U_{c}(\alpha )$ (the critical values
are represented by vertical lines in Fig. \ref{fig4}. At $U>U_{c}(\alpha )$,
the behavior is different, featuring $\left\langle n_{L/2}\right\rangle
=\left\langle n_{j}\right\rangle \equiv 1$. The latter regime means that the
inhomogeneous BHM, resembling its homogeneous counterpart \cite{fisher}, has
entered a state with the spatially uniform boson density, which is the
insulating Mott phase, where the superfluid density vanishes. 
It is well known 
that the Mott-superfluid transition in 1D is of the Kosterlitz-Thouless type
\cite{sachdev}, which means that the respective gap opens up exponentially,
hence precise determination of the transition point requires an accurate
finite-size scaling to extract the information pertaining to the
thermodynamic limit (the infinite system) \cite{kunen, white2, kollath}. As
Hamiltonian (\ref{ham}) is extensive, we cannot extrapolate to this limit;
nevertheless, a crude finite-size estimate of the transition point can be
worked out. In particular, as the repulsive interaction is weakest at the
two central sites, see Eq. (\ref{U}), one may conjecture that, once $U_{\min
}$ exceeds the critical value $U_{c}(\alpha =0)$, corresponding to the
Mott-superfluid transition in the homogeneous BHM, the transition to the
insulating Mott phase has occurred in the inhomogeneous system. From this
argument, the following value for the transition point can be extrapolated,
making use of Eq. (\ref{U}):
\begin{equation}
U_{c}(\alpha )=2^{\alpha }U_{c}(\alpha =0)~.  \label{analit}
\end{equation}%
Here we take the value of $U_{c}(\alpha =0)$ from Ref. \cite{kollath}, where
it was extrapolated to the thermodynamic limit. In Fig. \ref{fig4} the
vertical lines given by Eq. (\ref{analit}) are in qualitative agreement with
the numerical results. Moreover, as expected, the transition to the Mott
phase happens at smaller $U$ for lower $\alpha $.

\section{Conclusion}

In this work, we have introduced the 1D spatially inhomogeneous BHM
(Bose-Hubbard model) with the strength of the onsite self-repulsive
interaction growing from the center to periphery $\sim |j|^{\alpha }$, where
$j$ is the discrete coordinate. This model is the first one manifesting both
bright-soliton states and the insulating phase for purely repulsive
interactions. The model's Hamiltonian (\ref{ham}) is a quantum counterpart
of the semi-classical MF (mean-field) system, in the form of the DNLSE with
the strength of the onsite self-repulsion growing faster than $|j|$, which
gives rise to self-trapping of unstaggered localized modes \cite{boris}.
Here we have used the DMRG technique to construct the ground states of the
inhomogeneous BHM. In particular, in contrast with the previous results
produced by the MF system, we have shown that the quantum BHM gives rise to
weakly localized ground state at $\alpha <1$ just by tuning the strength of
the on-site interaction. Soliton-like self-trapped states have been found at
$\alpha >1$ for a broad range of the interaction strength, in agreement with
the MF limit. However, the essential difference from the MF counterpart is
that the soliton peak in the BHM is always surrounded by the background with
nonvanishing residual density. Eventually, still stronger repulsive
interactions destroy the soliton-like state, replacing it by the spatially
uniform Mott phase. An estimate for the critical interaction strength at the
transition point has been obtained.

To extend the present work, it may be interesting to construct higher-order
states (in particular, spatially odd modes), in the framework of both the
BHM and MF systems. A challenging issue is to extend the present analysis to
a two-dimensional inhomogeneous BHM.

\section*{Acknowledgments}

The authors acknowledge partial support from Università di Padova (grant No.
CPDA118083), Cariparo Foundation (Eccellenza grant 11/12), MIUR (PRIN grant
No. 2010LLKJBX). The visit of B.A.M. to Universitá di Padova was supported
by the Erasmus Mundus EDEN grant No. 2012-2626/001-001-EMA2. We thank M.
Dalmonte for useful discussions.


\begin{thebibliography}{99}
\bibitem{fisher} M. P. A. Fisher, G. Grinstein, and D. S. Fisher, Phys. Rev.
B \textbf{40}, 546-70 (1989).

\bibitem{bloch} M. Greiner, O. Mandel, T. Esslinger, T. Hänsch, and I.
Bloch, Nature \textbf{415}, 39 (2002).

\bibitem{bloch_review} I. Bloch, J. Dalibard, and W. Zwerger, Rev. Mod.
Phys. \textbf{80}, 885 (2008).

\bibitem{greiner} W. S. Bakr, J. I. Gillen, A. Peng, S. Foelling, M.
Greiner, Nature \textbf{462}, 74 (2009).

\bibitem{in-situ} J. F. Sherson, C. Weitenberg, M. Endres, M. Cheneau, I.
Bloch, S. Kuhr, Nature \textbf{467}, 68 (2010).

\bibitem{giamarchi} T. Giamarchi, Quantum Physics in One Dimension (Oxford
Univ. Press, Oxford, 2004).

\bibitem{cazalilla2011} M. A. Cazalilla, R. Citro, T. Giamarchi, E. Orignac,
and M. Rigol, Rev. Mod Phys. \textbf{83}, 1405 (2011).

\bibitem{quant-dark-sol} R. V. Mishmash, I. Danshita, C. W. Clark, and L. D.
Carr, Phys. Rev. A \textbf{80}, 053612 (2009); K. V. Krutitsky, J. Larson,
and M. Lewenstein, \textit{ibid}. \textbf{82}, 033618 (2010).

\bibitem{dark-sol} P. G. Kevrekidis, \textit{The Discrete Nonlinear Schrö%
dinger Equation: Mathematical Analysis, Numerical Computations, and Physical
Perspectives} (Springer: Berlin and Heidelberg, 2009).

\bibitem{Salerno} F. K. Abdullaev, B. B. Baizakov, S. A. Darmanyan, V. V.
Konotop, and M. Salerno, Phys. Rev. A\textbf{\ 64}, 043606 (2001); A.
Trombettoni and A. Smerzi, Phys. Rev. Lett. \textbf{86}, 2353 (2001).

\bibitem{salasnich} L. Barbiero and L. Salasnich, Phys. Rev. A \textbf{89},
063605 (2014).

\bibitem{exp-solo1} L. Khaykovich, F. Schreck, G. Ferrari, T. Bourdel, J.
Cubizolles, L. D. Carr, Y. Castin, and C. Salomon, Science \textbf{296},
1290 (2002).

\bibitem{exp-solo2} K. E. Strecker, G. B. Partridge, A. G. Truscott, and R.
G. Hulet, Nature (London) \textbf{417}, 150 (2002).

\bibitem{exp-solo3} S. L. Cornish, S. T. Thompson, and C. E. Wieman, Phys.
Rev. Lett. \textbf{96}, 170401 (3 May 2006).

\bibitem{exp-solo4} A. L. Marchant, T. P. Billam, T. P. Wiles, M. M. H. Yu,
S. A. Gardiner and S. L. Cornish, Nature Commun. \textbf{4}, 1865 (2013).

\bibitem{exp-solo5} J. H. V. Nguyen, P. Dyke, D. Luo, B. A. Malomed and R.
G. Hulet, arXiv:1407.5087v1

\bibitem{gap-sol} B. Eiermann, B. Th. Anker, M. Albiez, M. Taglieber, P.
Treutlein, K.-P. Marzlin, and M. K. Oberthaler, Phys. Rev. Lett. \textbf{92}%
, 230401 (2004).

\bibitem{morsch} O. Morsch and M. Oberthaler, Rev. Mod. Phys. \textbf{78},
179 (2006).

\bibitem{Barcelona0} O. V. Borovkova, Y. V. Kartashov, B. A. Malomed, and L.
Torner, Opt. Lett. \textbf{36}, 3088 (2011).

\bibitem{Barcelona} O. V. Borovkova, Y. V. Kartashov, L. Torner, and B. A.
Malomed, Phys. Rev. E \textbf{84}, 035602 (R) (2011).

\bibitem{boris} G. Gligori\'{c}, A. Maluckov, L. Hadzievski, and B. A.
Malomed, Phys. Rev. E \textbf{88}, 032905 (2013).

\bibitem{haus} Y. Lai and H. A. Haus, Phys. Rev. A \textbf{40}, 854 (1989).

\bibitem{drummond} P. D. Drummond, R. M. Shelby, S. R. Friberg, and Y.
Yamamoto, Nature (London) \textbf{365}, 307 (1993).

\bibitem{castin} Y. Castin and C. Herzog, Compt. Rendus Paris, serie IV,
\textbf{2}, 419 (2001).

\bibitem{weiss} C. Weiss and Y. Castin, Phys. Rev. Lett. \textbf{102},
010403 (2009).

\bibitem{weiss2} D. I. H. Holdaway, C. Weiss, and S. A. Gardiner, Phys. Rev.
A \textbf{85}, 053618 (2012).

\bibitem{delande} D. Delande, K. Sacha, M. Podzien, S. K. Avazbaev, and J.
Zakrzewski, New J. Phys. \textbf{15}, 045021 (2013).

\bibitem{weiss3} B. Gertjerenken, T. P. Billam, C. L. Blackley, C. R. Le
Sueur, L. Khaykovich, S. L. Cornish, and C. Weiss, Phys. Rev. Lett. \textbf{%
111}, 100406 (2013).

\bibitem{white} S. R. White, Phys. Rev. Lett. \textbf{69}, 2863 (1992).

\bibitem{book-lattice} M. Lewenstein, A. Sanpera, and V. Ahufinger, \textit{%
Ultracold Atoms in Optical Lattices: Simulating Quantum Many-Body Systems}
(Oxford University Press, Oxford, 2012).

\bibitem{book-bose} A. J. Leggett, Quantum Liquids. Bose condensation and
Cooper Pairing in Condensed-Matter Systems (Oxford University Press, Oxford,
2006).

\bibitem{nlattice} Y. V. Kartashov, B. A. Malomed, and L. Torner, Rev. Mod.
Phys. \textbf{83}, 247 (2011).

\bibitem{Dong} T. Mayteevarunyoo, B. A. Malomed, and G. Dong, Phys. Rev. A
\textbf{78}, 053601 (2008).

\bibitem{Roati} G. Roati, M. Zaccanti, C. D'Errico, J. Catani, M. Modugno,
A. Simoni, M. Inguscio, and G. Modugno, Phys. Rev. Lett \textbf{99}, 010403
(2007).

\bibitem{Pollack} S. E. Pollack, D. Dries, M. Junker, Y. P. Chen, T. A.
Corcovilos, and R. G. Hulet, Phys. Rev. Lett \textbf{102}, 090402 (2009).

\bibitem{Boris1} B. A. Malomed, D. Mihalache, F. Wise, and L. Torner, J.
Optics B: Quant. Semicl. Opt. \textbf{7}, R53 (2005).

\bibitem{Gora} P. O. Fedichev, Yu. Kagan, G. V. Shlyapnikov, and J. T. M.
Walraven Phys. Rev. Lett \textbf{77}, 2913 (1996).

\bibitem{Durr} D. M. Bauer, M. Lettner, C. Vo, G. Rempe, and S. Dürr, Nature
Phys. \textbf{5}, 339 (2009)

\bibitem{Boshier} K. Henderson, C. Ryu, C. MacCormick, and M. G. Boshier,
New J. Phys. \textbf{11}, 043030 (2009)

\bibitem{note} Mean-field results obtained from the DNLSE are fully reliable
only when $U\rightarrow 0$ and $N\rightarrow \infty $ with $UN$ taken
constant, see Ref. \cite{salasnich} and references therein.

\bibitem{book-luca} L. Salasnich, Quantum Physics of Light and Matter: A
Modern Introduction to Photons, Atoms and Many-Body Systems (Springer,
Berlin, 2014).

\bibitem{sala-numerics} E. Cerboneschi, R. Mannella, E. Arimondo, and L.
Salasnich, Phys. Lett. A \textbf{249}, 495 (1998); G. Mazzarella and L.
Salasnich, Phys. Lett. A \textbf{373}, 4434 (2009).

\bibitem{manmana} C. P. Rubbo I. I. Satija, W. P. Reinhardt, R.
Balakrishnan, A. M. Rey and S. R. Manmana, Phys. Rev. A \textbf{85}, 053617
(2012).

\bibitem{das} T. Mishra, R. V. Pai, S. Ramanan, M. S. Luthra and B. P. Das,
Phys. Rev. A \textbf{80}, 043614 (2009).

\bibitem{sacha} D. Delande, K. Sacha, M. Plodzien, S. K. Avazbaev, and J.
Zakrzewski, New J. Phys. \textbf{15} 045021 (2013).

\bibitem{magnons} T. Fukuhara, P. Schauss, M. Endress, S. Hild, M. Cheneau,
I. Bloch, and C. Gross, Nature \textbf{502}, 76 (2013)

\bibitem{comment1} Of course this equality is true only for even $L$, as in
the present case.

\bibitem{sachdev} S. Sachdev, \textit{Quantum Phase Transitions} (Cambridge
University Press, 1999).

\bibitem{kunen} T. D. Kuhner and H. Monien, Phys. Rev. B \textbf{58}, R14741
(1998).

\bibitem{white2} T. D. Kuhner, S. R. White and H. Monien, Phys. Rev. B
\textbf{61}, 18 (2000).

\bibitem{kollath} A. L\"auchli and C. Kollath, J. Stat. Mech., P05018 (2008).

\end{thebibliography}
\end{document}